\documentclass{article}
\usepackage[accepted]{vietnam} 
\usepackage{natbib}
\usepackage{graphicx}      
\usepackage{amsmath, amssymb, amsthm}

\usepackage{color}
\usepackage{hyperref} %----set up hyper link to citations, equations...
\hypersetup{colorlinks=true,citecolor=blue}

\begin{document}
\twocolumn[
\title{Dust Growth and Magnetic Fields:\\ from Cores to Disks (even down to Planets)}
\titlerunning{Dust Growth and Magnetic Fields}
\author{Yasuhiro Hasegawa}{yasuhiro@caltech.edu}
\address{Jet Propulsion Laboratory, California Institute of Technology, Pasadena, CA 91109, USA}
%if there are more authors, just add more. Otherwise, please comment it.
\author{Jennifer I-Hsiu Li}{jli184@illinois.edu}
\address{Department of Astronomy, University of Illinois at Urbana-Champaign, Urbana, IL 61801, USA}
\author{Satoshi Okuzumi}{okuzumi@geo.titech.ac.jp}
\address{Department of Earth and Planetary Sciences, Tokyo Institute of Technology, Meguro-ku, Tokyo 152-8551, Japan}

\vskip 0.5cm 
]

\begin{abstract}
The recent rapid progress in observations of circumstellar disks and extrasolar planets has reinforced 
the importance of understanding an intimate coupling between star and planet formation.
Under such a circumstance, 
it may be invaluable to attempt to specify when and how  planet formation begins in star-forming regions
and to identify what physical processes/quantities are the most significant to make a link between star and planet formation.
To this end, we have recently developed a couple of projects.
These include an observational project about dust growth in Class 0 YSOs and a theoretical modeling project of the HL Tauri disk.
For the first project, we utilize the archive data of radio interferometric observations, 
and examine whether dust growth, a first step of planet formation, occurs in Class 0 YSOs. 
We find that while our observational results can be reproduced by the presence of large ($\sim$ mm) dust grains for some of YSOs 
under the single-component modified blackbody formalism,
an interpretation of no dust growth would be possible when a more detailed model is used.
For the second project, we consider an origin of the disk configuration around HL Tauri, 
focusing on magnetic fields. 
We find that magnetically induced disk winds may play an important role in the HL Tauri disk.
The combination of these attempts may enable us to move towards a comprehensive understanding 
of how star and planet formation are intimately coupled with each other.
\end{abstract}

\section{Introduction: Implications from the recent astonishing observational discoveries}

Star formation is one of the most crucial processes in the Universe.
This is because star formation can influence the evolution of the Universe over a wide range of physical scales. 
For instance, the presence of stars can affect the formation and evolution of galaxies through the so-called stellar feedback.
Another example may be that star formation can presumably trigger the subsequent formation of planetary systems.

It has long been suggested that star and planet formation are intimately coupled with each other (see \citealt{hayashi81}).
This coupling possibly occurs via the formation and evolution of circumstellar disks 
that emerge as a consequence of the conservation of angular momentum in a star formation stage (\citealt{li14}).
The disks can also serve as the birth place of planetary systems in the subsequent planet formation stage (\citealt{benz14}).
It is paramount that the recent, unprecedented high spatial resolution ALMA observations taken toward a young stellar object (YSO), HL Tauri,
have partially confirmed such a coupling between stars and planets (\citealt{alma15});
the observations reveal that nearly concentric multiple gaps are present in the observed dust continuum emission originating from the disk.
While an ultimate origin of such gaps is still under active investigation,
one of the most intriguing hypotheses is that unseen, forming (proto)planets would be responsible for the formation of the gaps (e.g., \citealt{akiyama16}).
Furthermore, the recent rapid accumulation of extrasolar planets, also known as exoplanets, has suggested that 
planetary systems around the main-sequence stars are ubiquitous in the Universe (\citealt{winn15}).
This observational result also provides additional supportive evidence that there is an intimate coupling between star and planet formation.

Based on these recent astonishing observational discoveries,
it is in almost no doubt that planet formation is a natural outcome of star formation.
It is, however, important to point out that these observations would readily pose the following questions:
when does (a first step of) planet formation occur in the context of star formation?
And what processes/quantities would be ideal to explore an intimate coupling between star and planet formation?
These questions are interesting because HL Tauri is currently regarded as a quite young ($\lesssim 1$ Myr), Class I-II YSO.
In other words, if the observed gaps would be caused by unseen planets,
then the ALMA observations suggest that planet formation should have taken place much earlier than expected before.
And if this would be the case, there might be a possibility that 
one can trace the formation history of planets back to the initial condition of star formation,
by focusing on certain physical processes/quantities.

To address these questions, we have developed a couple of projects,
one of which is to observationally investigate dust growth in Class 0 YSOs (\citealt{li17}),
the other of which is to theoretically model the circumstellar disk around HL Tauri, 
with the emphasis on magnetic fields threading the disk (\citealt{hasegawa17}).
We will discuss below that dust growth and magnetic fields are the important ingredients 
to examine a vital connection between star and planet formation.

\section{An observational study about dust growth in Class 0 YSOs}
\begin{table*}
\caption{List of our targets and the resultant value of $\beta$}
\centering  
\label{table1} 
\begin{tabular}{llclc}
\hline
Source Name             &  Region    & Distance (pc)   & Class     & Best Fit ($\beta$)  \\ 
\hline
L1527                          & Taurus     & 140                  & 0/I          & 0.30         \\
L1448CN 			  & Perseus   & 240                  & 0            & 0.58         \\
NGC1333 IRAS4 A1   & Perseus   & 240                  & 0            & 0.51          \\
NGC1333 IRAS4 A2   & Perseus   & 240                  & 0            & 1.23          \\
NGC1333 IRAS2A      & Perseus   & 240                  & 0            & 1.20          \\
Barnard 1b N               & Perseus  & 240                  & 0            &  1.53         \\
Barnard 1b S               & Perseus   & 240                 & 0            & 2.02          \\
L1157                          & Isolated    & 325                 & 0            & 0.82          \\
Serpens FIRS 1          & Serpens  & 415                  & 0/I          & 1.74         \\
\hline
%\bottomrule
\end{tabular}
\end{table*}

\begin{figure*}
\vskip -0.5cm
\centering
%$\begin{array}{cc}
\includegraphics[angle=0,width=11cm]{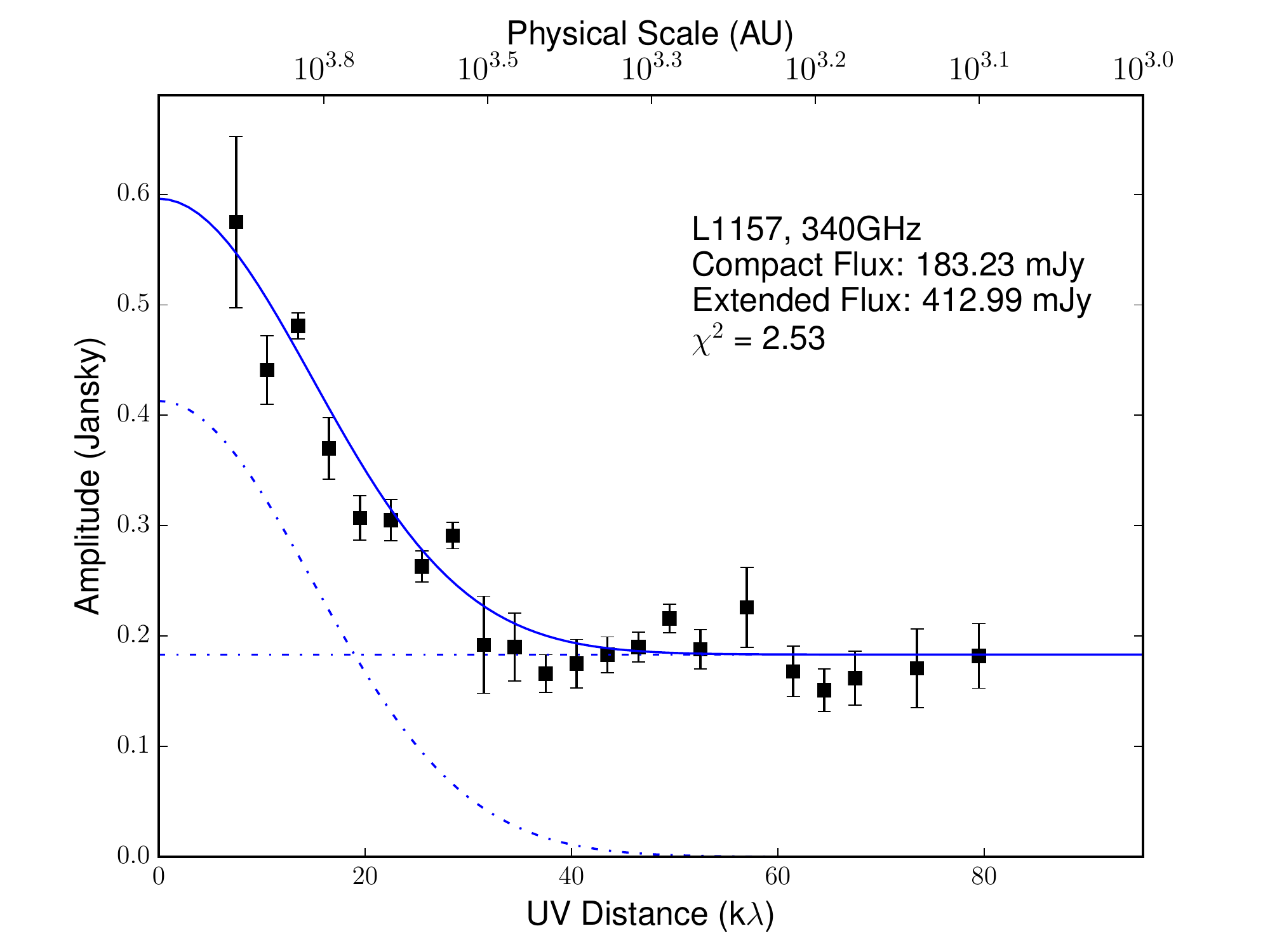} 
\includegraphics[angle=0,width=10cm]{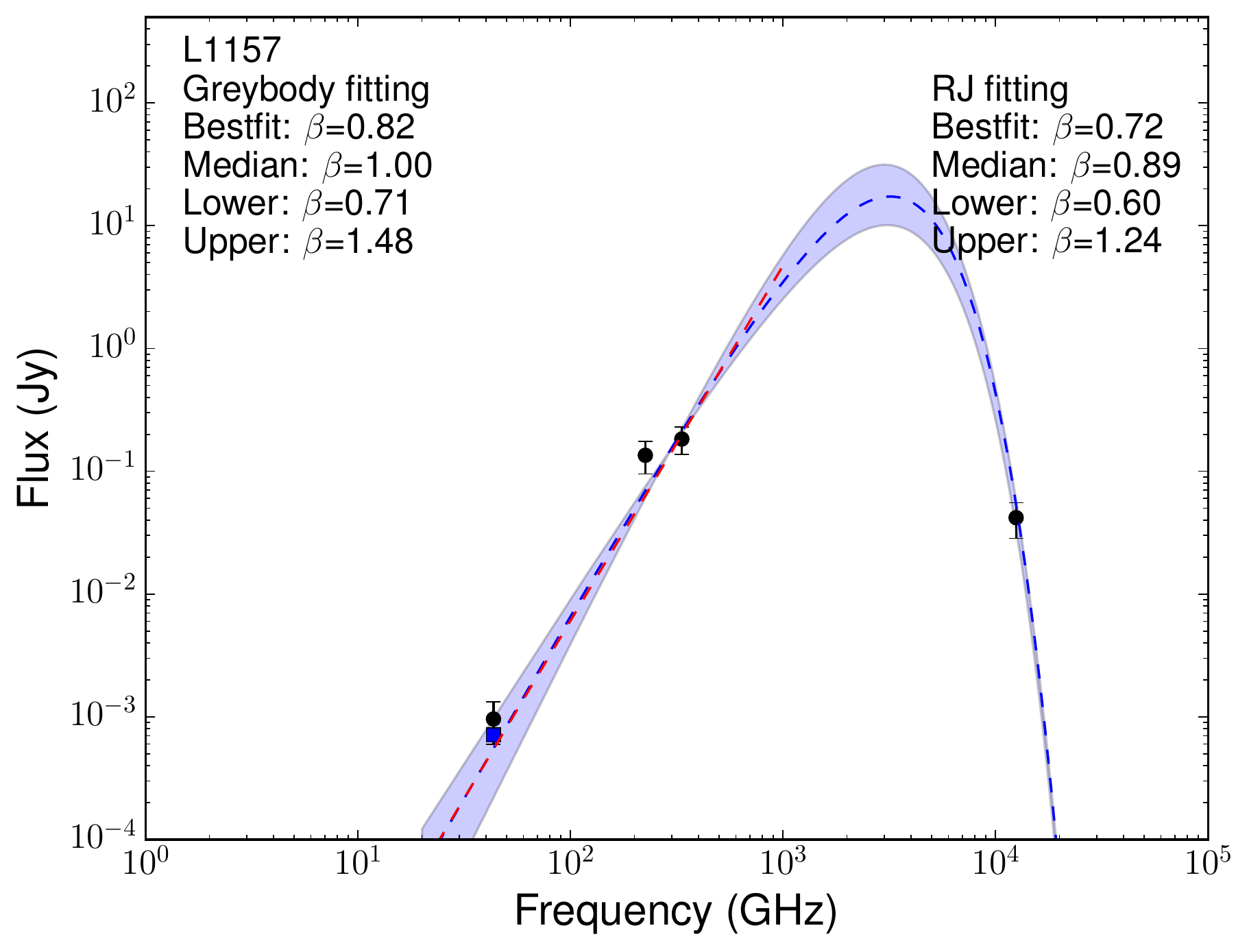} 
%\end{array}$
\caption{An example of the visibility data (top) and the result of SED fitting (bottom) for L1157.}
\label{fig1}
\vspace{-0.5cm}
\end{figure*}

Here, we introduce our project about dust growth in Class 0 YSOs (\citealt{li17}).
The main motivation of this project is to identify when dust growth, a first step of planet formation, occurs in an early stage of star formation.

In order to proceed, we utilize radio, interferometric observations covering submillimeter to cm wavelengths,
and perform spectral energy distribution (SED) fitting based on a modified black-body radiation formalism.
Longer wavelength observations are required to examine whether large ($\sim$ 1mm) dust grains are present 
in the high density, star-forming region.
While collisional growth of dust grains can be possible in such a high density region,
the thermal emission from large dust grains becomes optically thin(ner) only at submillmeter wavelengths or longer there.
In principle, at least 3 different frequency band observations are  needed in SED fitting
to simultaneously determine the fundamental properties of dust grains such as the dust mass, the dust size, and the dust temperature.
We make use of the abundant archival data in order to collect a good quality of data.
More specifically, we retrieve the archive data of (J)VLA, SMA, CARMA, and ALMA. 

In this project, we attempt to undertake a survey type of studies.
This is because
the previous studies reported a discrepancy between observations and theory;
the previous (sub)millimeter observations suggest that 
large ($\sim$ 1mm) dust grains would be needed to fully reproduce the observed value of the mm opacity spectral index
(the so-called $\beta$) for some of class 0/I YSOs (e.g., \citealt{jorgensen07,ricci10}).
Theoretical studies, however, indicate that the presence of such large grains would be very unlikely in these systems
(e.g., \citealt{ormel09}; \citealt{hirashita13}).
This is simply because these systems are in general short-lived and a relatively low density,
both of which prevent collisional growth of dust particles.
Thus, it is important to systematically investigate a large number of targets
to address whether large dust grains exist in Class 0 YSOs.

Based on our close retrieval of the rich archive data, 
we find that 9 well-studied Class 0 YSOs can meet the purpose of our SED fitting (see Table \ref{table1}).
All the targets are chosen from nearby star-forming regions due to the (low) angular resolution of SMA archive data.
In addition to the requirement that 3 different wavelength observations should be available for SED fitting (see above),
we request that the selected targets exhibit at least one significant detection at $>$ 3 mm wavelength observations
where the dust thermal emission becomes optically thinner.
Furthermore, Spitzer 24 and 27 $\mu$m MIPS data are also included
in order to cover a high frequency wing in the SED plot.

Once a list of our targets are complied and the standard data reduction is performed,
we develop and carry out a systematic data analysis procedure 
to derive the dust properties in Class 0 YSOs from SED fitting.
This procedure consists of three steps.
The first step is to dynamically bin the azimuthally-averaged visibility amplitude as a function of $uv$ distance  
to properly weight a collection of observations that were conducted in different weather conditions and for different science purposes.
This automatized binning allows us to obtain a good quality of data that can have a good value of the signal-to-noise ratio (SNR)
while keeping the spatial information.
The second step is to decompose the resultant visibility amplitude into two components,
one of which is a spatially compact, disk-like component,
the other of which is a spatially extended, envelope component.
This decomposition is motivated by a current picture of star formation (e.g., \citealt{li14}).
The final step is to perform SED fitting using a modified black body formalism.
Based on our preliminary results, only the compact component can be utilized for our SED fitting due to its good data quality.

Figure \ref{fig1} shows an example of our results for L1157.
The top panel depicts the automatically binned, visibility amplitude as a function of $uv$ distance,
and the bottom one is for the result of our SED fitting.
We find that the compact component for this target is well represented by a straight line 
where the visibility amplitude is constant (the top panel).
This indicates that the compact, disk-like structure is not resolved by the observations at all, and is regarded as a point source.
On the other hand, the extended component is well fitted by a Gaussian profile, and is viewed as a blob in the image domain.
For the SED fitting (the bottom panel),
our results show that a best fit is obtained 
when the resultant value of $\beta$ is about 0.8 for the compact component.
This suggests that large ($\sim$ mm) dust grains should be present in this target.
In order to verify whether or not this result would be applicable for other YSOs,
we analyze the rest of our targets using the same procedure.
Our results are summarized in Table \ref{table1} (see the Best Fit ($\beta$)).
One immediately notices that dust growth (a low value of $\beta(<1)$) very likely takes place in roughly half of our targets,
which supports the previous observational results.
On the contrary, our results suggest that the other half of our targets have a high value of $\beta(>1)$,
which tends to be consistent with the results of theoretical studies.
Thus, our SED fitting implies that the degree of grain growth may depend on the properties of YSOs,
which may be different for different targets.

How are our results and interpretations reasonable?
One of the advantages of our method is its simplicity,
which permits a systematic analysis of the archive data for a number of targets.
It is nonetheless obvious that such simplicity is established by imposing a number of (implicit) assumptions and approximations.
One of the most serious assumptions is that the dust properties (the dust mass, the dust size, and the dust temperature)
are all represented by single quantities for the compact component.
In other words, there is no spatial information for them.
In fact, our derived value of $\beta$ should be regarded as the mean value for the compact component.
It is important to realize that 
all the dust quantities are indeed a function of the distance from the central source.
As an example, the dust temperature would be much higher in the inner part of the compact component than the outer part.
Furthermore, different wavelength observations generally detect the emission originating from different regions due to the optical depth effect.
Thus, much higher spatial resolution observations, fine-spaced SED sampling, and/or more detailed radiative transfer modeling 
would be demanded to fully address the presence of large ($\sim$ mm) dust grains in Class 0 YSOs.
In addition, it would be useful to investigate this using another method such as a chemistry model (\citealt{harada17}).
A more complete SED modeling and discussion can be found in \citealt{li17}.

\begin{figure*}[!ht]
%\vskip -0.5cm
\centering
$\begin{array}{cc}
\includegraphics[angle=0,width=8.5cm]{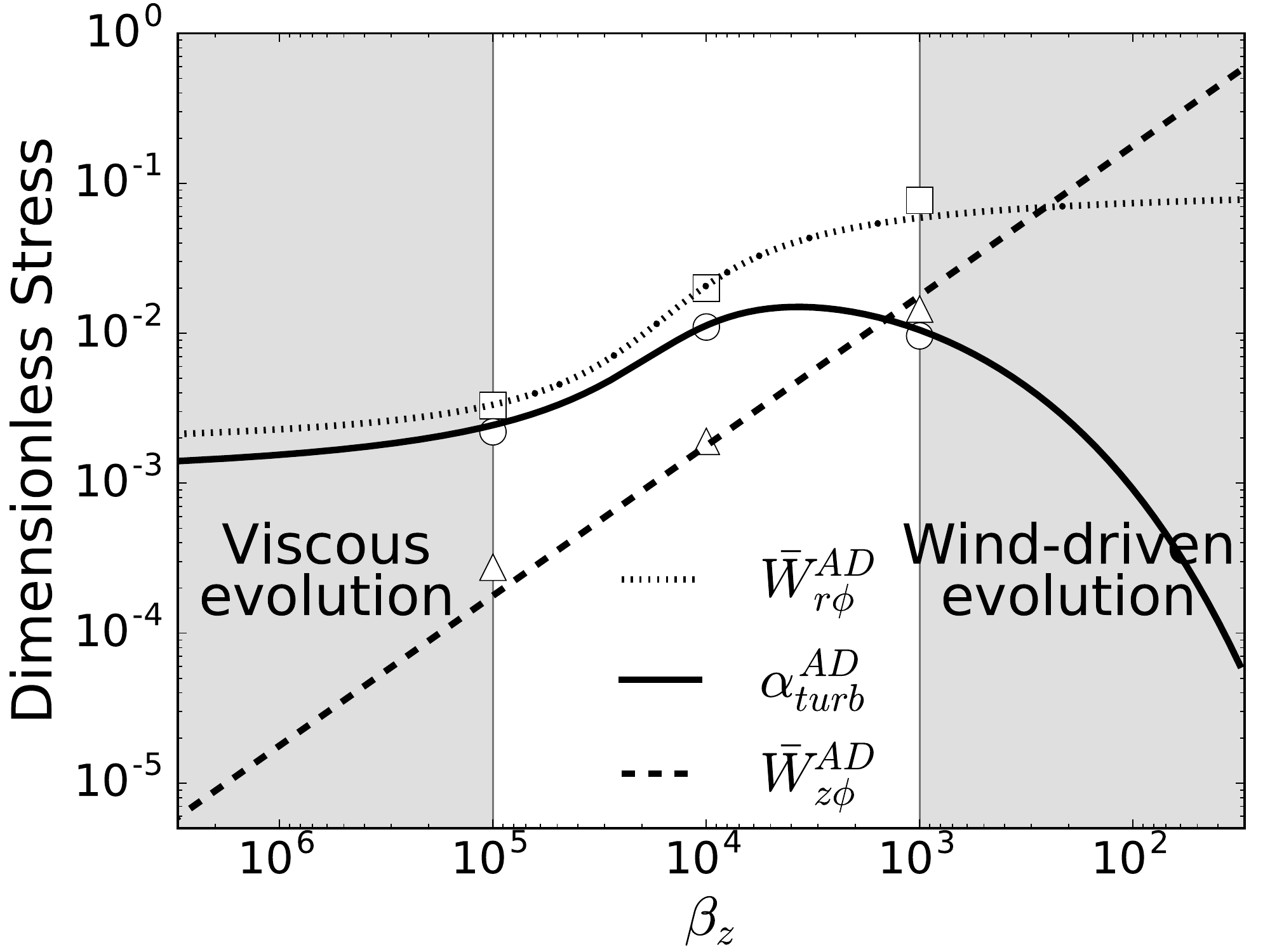} 
\includegraphics[angle=0,width=8.5cm]{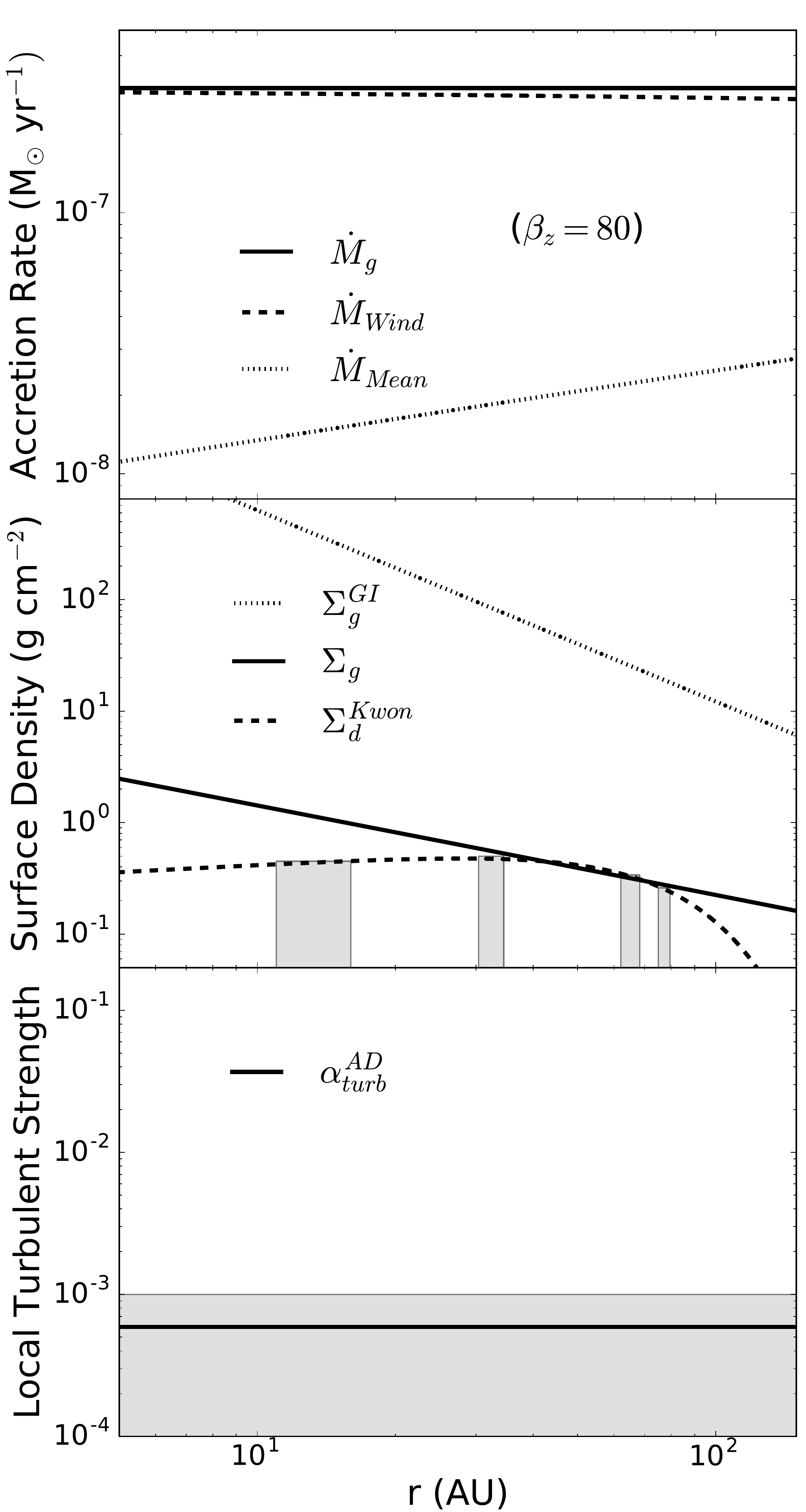} 
\end{array}$
\caption{Our semi-analytical formulae for the accretion stress ($\bar{W}^{\rm AD}_{r\phi}$), 
the wind stress ($\bar{W}^{\rm AD}_{z\phi}$), and the local component of disk turbulence ($\alpha^{\rm AD}_{\rm turb}$) as a function of the plasma $\beta_z$ (left),
and the resultant disk properties for the case that $\beta_z =80$ (right).
On the right panel, the disk accretion rate, the gas surface density profile, and the value of $\alpha_{\rm turb}$ are plotted as a function of the distance from the central star,
from top to bottom.}
\label{fig2}
%\vspace{-0.5cm}
\end{figure*}

\section{A theoretical modeling of the magnetized circumstellar disk around HL Tauri}

We now discuss another project in which a theoretical modeling of the circumstellar disk around HL Tauri is developed (\citealt{hasegawa17}).
The main motivation of this project is to demonstrate how magnetic fields threading disks would be useful 
to make a vital connection between star and planet formation.
 
As described above, the ALMA observations of HL Tauri are revolutionizing a current picture of star and planet formation.
While one of the most striking features in the observations is the nearly concentric multiple gaps in the dust continuum emission,
we here focus on the interplay between magnetically driven disk accretion and the vertical distribution of dust particles in the disk.

It has been suggested that HL Tauri can probably be classified as a young $(\lesssim 1$ Myr), Class I-II YSO.
In fact, the previous observations taken toward HL Tauri reveal a number of characteristic features of YSOs
such as molecular outflow and the presence of an envelope (\citealt[references herein]{akiyama16}).
Furthermore, some observations and modeling indicate that 
the disk accretion rate onto the central star is about $10^{-7}-10^{-6}$ M$_{\odot}$ yr$^{-1}$ (\citealt{hayashi93,beck10}).
When the famous $\alpha-$prescription is adopted (\citealt{shakura73}),
this accretion rate corresponds to $\alpha_{\rm acc}$ which is $10^{-2}-10^{-1}$.
Note that $\alpha$ is labeled as $\alpha_{\rm acc}$ to explicitly denote that this $\alpha$ is derived from the disk accretion rate.
Such a high value of $\alpha_{\rm acc}$ may be reasonable for HL Tauri due to its young age.
At the same time, the recent ALMA observations provide another opportunity to conduct a detailed radiative transfer modeling.
\citet{pinte16} have undertaken this and found that the vertical distribution of dust particles is very thin ($\lesssim 1$ AU at the semimajor axis of $\sim$ 100 AU).
Based on their modeling, 
the required value of $\alpha$ (defined as $\alpha_{\rm turb}$) should be as small as a few $10^{-4}$
in order to realize such a high degree of vertical dust settling.
When turbulence is the main driver of disk accretion, it is generally expected that $\alpha_{\rm acc} \simeq \alpha_{\rm turb}$,
since most of disk turbulence would be more or less local processes.
For the HL Tauri case, however,
it is obvious that there is more than one order magnitude difference between $\alpha_{\rm acc}$ and $\alpha_{\rm turb}$.

In order to reconcile this big mismatch,
we develop a simple, but physically motivated disk model.
In this model, it is assumed that 
disk accretion onto the central star is driven not only by local turbulence (which is adopted in many of disk models),
but also by magnetically induced disk winds.
The recent numerical simulations show that disk winds can be launched at disk surfaces and remove a substantial amount of the disk angular momentum vertically 
under the condition that disks are threaded by relatively strong, vertical magnetic fields  (e.g., \citealt{suzuki09,simon13}).
Disk winds would play an important role in the evolution especially when ambipolar diffusion, a non-ideal MHD effect, is properly taken into account (e.g., \citealt{bai11}).
Practically, we first develop semi-analytical formulae for the accretion stress ($\bar{W}^{\rm AD}_{r\phi}$), 
the wind stress ($\bar{W}^{\rm AD}_{z\phi}$), and the local component of disk turbulence ($\alpha^{\rm AD}_{\rm turb}$),
as a function of the plasma $\beta_z$.
Note that $\bar{W}^{\rm AD}_{r\phi} \propto \alpha_{acc}$.
Also, beware of $\beta_z$ which is different from the opacity index ($\beta$) that appeared in Section 2.
These formulae are obtained, so that they can fit the numerical simulations very well (\citealt[][see the left panel of Figure \ref{fig2}]{simon13}).
We then parameterize the plasma $\beta_z$ to find out a disk configuration that can account for the HL Tauri disk,
under the assumption of steady disk accretion.

Figure \ref{fig2} (the right panel) shows the resultant disk properties for the case that $\beta_z = 80$.
From top to bottom, the disk accretion rate, the gas surface density profile, and the value of $\alpha_{\rm turb}^{\rm AD}$ are plotted 
as a function of the distance from the central star.
As a reference, the dust surface density distribution (denoted by $\Sigma_d^{\rm Kwon}$) that is derived from CARMA observations (\citealt{kwon11}),
and the locations of gaps (denoted by the hatched regions) detected by ALMA observations (\citealt{alma15}) are both included in the middle.
Our results show that when magnetically induced disk winds are taken into account,
the winds contribute predominantly to the disk accretion rate (see the top).
This occurs because the wind stress ($\bar{W}^{\rm AD}_{z\phi}$) can be scaled as $\alpha_{acc} r / h$,
where $r$ is the distance from the central star, $h$ is the pressure scale height, and $r/h \gg 1$.
For the gas distribution (the middle),
we find that the gas surface density becomes relatively low.
This is a direct outcome that the wind can remove the disk angular momentum significantly.
It is interesting that the resultant value of the gas surface density may be comparable to that of dust,
especially at the locations of observed gaps.
Finally, our results (the bottom) indicate that disk wind models can reproduce a low value of $\alpha_{\rm turb}$
that is suggested for the HL Tauri disk (see the hatched region).
This becomes possible thanks to the winds that can establish efficient disk accretion onto the central star
without a high level of disk turbulence.
Thus, our modeling suggests that magnetically induced disk winds can achieve a disk configuration
where the disk accretion rate is high due to the winds
while the local turbulence is weak enough to realize a high degree of vertical dust setting.

How can our current, simplified modeling be improved?
There might be at least 3 ways to sophisticate our model.
First, one can adopt more realistic formulae for $\bar{W}^{\rm AD}_{r\phi}$, $\bar{W}^{\rm AD}_{z\phi}$, and $\alpha^{\rm AD}_{\rm turb}$,
following the more recent work such as \citet{bai16,suzuki16}.
Second, the vertical distribution of dust particles can be computed directly 
from the dust size and magnetically driven disk turbulence, for a given value of the plasma $\beta_z$ (e.g., \citealt{zhu15}).
Third, a global, time-dependent, 1D disk model can be developed
by implementing the above two improvements.
Such a disk model would be useful to examine how magnetic fields are related to the observed multiple gaps in the HL Tauri disk,
and hence ultimately serve as a powerful tool to explore an intimate coupling between star and planet formation.
A more complete discussion can be found in \citealt{hasegawa17},
where some improvements are already undertaken.

\section{Summary and future prospects}

We have discussed two projects that have been developed to address when a first step of planet formation occurs in star-forming environments
and what processes/quantities are useful to make a vital link between star and planet formation.
As described above, our current projects can provide only a partial answer to these questions.
In our future work,  we will undertake much higher spatial resolution observations and/or more detailed modeling 
to obtain a more comprehensive understanding 
of how star and planet formation take place together and of how these two processes are intimately coupled with each other.

\section*{Acknowledgments}
%\vspace{-0.3cm}

This research was carried out at Jet Propulsion Laboratory, California Institute of Technology under a contract with NASA.
Y.H. is supported by JPL/Caltech.

%\section*{References}

\end{document}